\documentclass[conference]{IEEEtran}

\IEEEoverridecommandlockouts
\usepackage{etex}
\usepackage{xcolor}
\usepackage{cite}
\usepackage[numbers,sort&compress]{natbib}

\usepackage{amsmath,amssymb,amsfonts}
\usepackage{graphicx}
\usepackage{subcaption}
\usepackage{textcomp}
\usepackage{amsmath}
\usepackage{xcolor}
\usepackage{caption}
\usepackage{float}

\usepackage{algorithm} 
\usepackage{algorithmic}  
\usepackage[algo2e]{algorithm2e} 
\usepackage{url}
\usepackage{array}
\usepackage{booktabs}
\usepackage{hhline}
\graphicspath{ {./fig/} }
\usepackage[section]{placeins}
\usepackage{mathtools}
\usepackage{multirow}

\graphicspath{ {./figures/} }
\def\BibTeX{{\rm B\kern-.05em{\sc i\kern-.025em b}\kern-.08em
    T\kern-.2em\lower.7ex\hbox{E}\kern-.125emX}}

\begin{document}

\title{ECO-ID: Event-Camera based Optical System for Secure Multi-User Ultra-Low Latency Identification}

\author{
\IEEEauthorblockN{Subham Sabud, Chengling Xu, and Feng Ye
\IEEEauthorblockA{Department of Electrical and Computer Engineering, University of Wisconsin-Madison, Madison, WI, USA\\
Emails: \IEEEauthorrefmark{1}\{sabud, cxu338, feng.ye\}@wisc.edu
}
}
}

\maketitle

\begin{abstract}
Time-critical interactive systems increasingly require ultra-low-latency device identification for multiple users, yet prevailing approaches such as passwords, QR codes, RFID/NFC are constrained by human input, frame-based sensing, or near-contact range. This paper presents \textbf{ECO-ID}, an event-camera-based optical system for multi-user, ultra-low-latency identification over visible light communication (VLC). Leveraging microsecond-resolution, asynchronous observations of brightness transitions, ECO-ID employs a spatiotemporal coding design: disjoint LED subsets provide spatial separation among users, while user-specific timing delays encode identities without inter-user synchronization. The optical channel and event-driven sensing reduce full-scene capture relative to frame cameras and limit the RF attack surface, while enabling rapid token verification with freshness and replay protection. We implement a prototype and demonstrate that ECO-ID can practically achieve approximately 99.8\% localization and 98.7\% identification with 0.64 ms mean latency, while theoretically supporting identification at the scale of tens of concurrent users. Overall, ECO-ID provides a fast, privacy-conscious, and security-aware alternative for scalable multi-user identification in time-critical interactive environments.
\end{abstract}

\section{Introduction}

Device authentication is fundamental to modern digital systems, ensuring that only authorized users can access devices and services. In emerging applications, authentication is increasingly a time-critical primitive. For example, multi-user AR/VR, time-sensitive human-machine interaction, and intelligent cyber-physical systems must rapidly determine who is present and authorized, often for multiple users concurrently. Despite its importance, however, ultra-low-latency identification remains difficult to achieve reliably with today’s dominant technologies.
Most deployed identification mechanisms rely on one-time interaction, such as QR codes, RFID/NFC tokens, or pre-shared credentials~\cite{ogorman2003comparing,bonneau2012quest}. They are limited by human input, frame-based sensing latency, or near-contact range: passwords incur reaction time and usability costs; QR-code scanning is bounded by frame acquisition/decoding and sensitivity to blur and illumination~\cite{vangennip2015qr,sorgel2015fastblur}; and RFID/NFC typically requires centimeter-scale proximity~\cite{want2006rfid,coskun2013nfcsurvey} while remaining vulnerable to relay and eavesdropping attacks~\cite{francis2010relay}. These constraints also complicate secure communication: visual tokens can be copied and replayed, RF links can be proxied, and frame-based cameras may capture sensitive contextual information.

To address these limitations, we propose \textbf{ECO-ID}, an event-camera-based optical system for multi-user, ultra-low-latency identification. ECO-ID combines visible light communication (VLC)~\cite{karunatilaka2015vlc,matheus2019vlc} with an event camera that reports asynchronous brightness changes~\cite{lichtsteiner2008dvs,gallego2022event}. The sensor outputs sparse events with microsecond-level timestamps only when intensity changes occur, enabling rapid detection of LED switching transitions while reducing scene capture relative to conventional imaging. ECO-ID uses spatiotemporal coding for concurrent, asynchronous users: disjoint LED subsets provide spatial separation; a short localization pattern produces a coincident event burst to mark onset; and an ID-LED activation after a user-specific delay encodes identity. The receiver aggregates events in per-LED ROIs, detects each user’s onset, and estimates identity from the relative delay to the ID LED, enabling multi-user association without inter-user synchronization or scheduling.

Importantly, ECO-ID also strengthens the security posture of the identification communication itself. Compared to RF-based token exchange, directional optical signaling reduces unintended exposure beyond the line-of-sight (LoS) region, and event-driven reception avoids recording full frames of the environment. ECO-ID further supports secure token verification by incorporating authenticated tags together with freshness checks (timestamps/TTL) and nonce-based replay protection, mitigating token reuse even when observations are attempted by an adversary. Compared to QR-code scanning, ECO-ID is robust to imperfect focus because it relies on temporal brightness transitions rather than frame-based features, and it supports room-scale, fully contactless operation under a LoS path.
ECO-ID further strengthens the identification communication. Directional optical signaling and ultra-high temporal resolution of ECO-ID limits RF and spectral-camera style attacks. The ultra-low latency and timestamp based processing can straightforwardly incorporate authenticated tags with freshness (timestamps/TTL) and nonce-based replay protection, mitigating token reuse under observation. ECO-ID also supports room-scale, contactless operation under LoS and remains robust to imperfect focus by relying on temporal transitions rather than frame features.

We implement a prototype of ECO-ID and evaluate localization success, identification success, and end-to-end identification latency over 2-30~kHz switching frequencies, including asynchronous and overlapping transmissions. Results show sub-millisecond identification across all tested frequencies; at 4~kHz, ECO-ID achieves $\approx$99.8\% localization and $\approx$98.7\% identification with $\sim$0.64~ms mean latency for both users, and supports identification at the scale of tens of concurrent users. The main contributions of this work are threefold: (i) ECO-ID, an event-camera VLC identification/authentication framework for time-critical interactive environments; (ii) a spatial-temporal coding protocol with disjoint LED subsets and timing signatures enabling concurrent multi-user identification with $\mu$s-level transition detection, sub-ms latency, and meter-scale LoS operation; and (iii) prototype implementation and experimental validation demonstrating sub-ms identification for multiple devices.

\section{System Model and Prototype}\label{sec:system model}

\begin{figure}[ht!]
    \centering
    \includegraphics[width=0.99\linewidth]{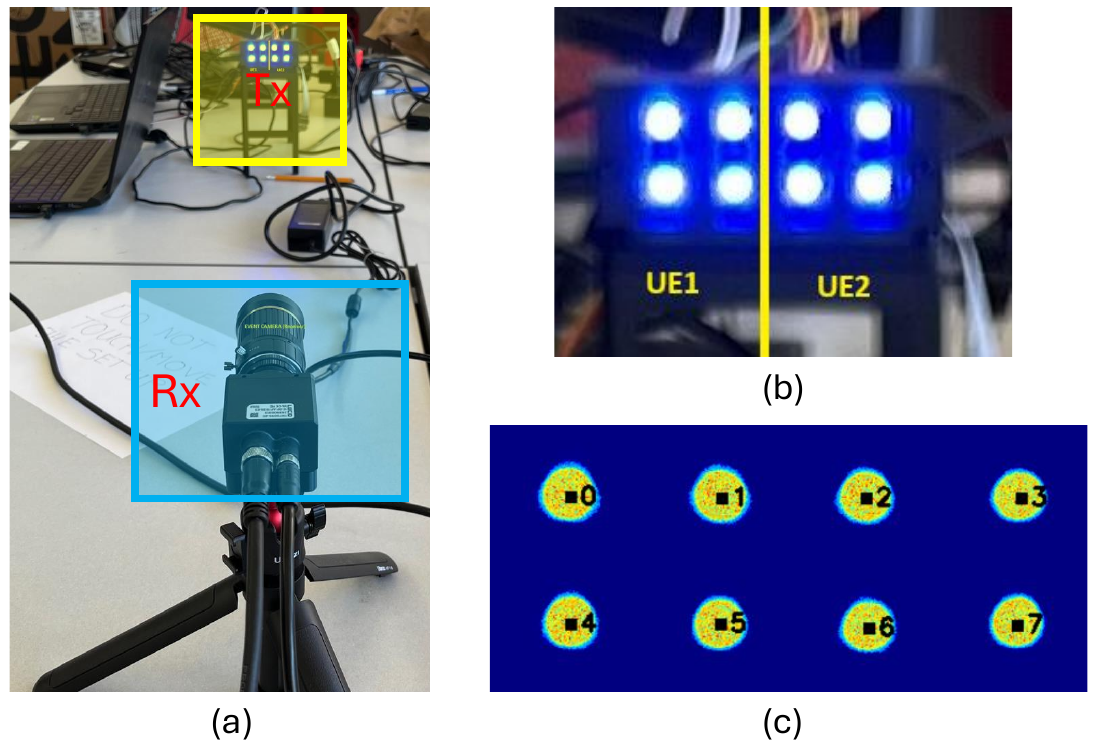}
    \caption{Testbed setup: (a) Overall system in action; (b) Two-user setting (UE1 on the left, and UE2 on the right); and (c) Example of out-of-focus LED observations captured by the event camera. A $9 \times 9$ pixel ROI is selected around the center of each LED for robust event detection.}
    \label{fig:setup_led_panel}
\end{figure}

We consider a VLC system in which spatially distributed light-emitting elements act as transmitters and an event-based vision sensor acts as the receiver. The transmitter comprises one or more LED panels, each containing multiple independently controllable LEDs. Let $\mathcal{L}$ denote the set of all LEDs across panels, with $|\mathcal{L}|=L$. Each LED $\ell\in\mathcal{L}$ is driven by an on/off control signal to induce brightness transitions detectable by the receiver. For concreteness, we instantiate the model using the prototype shown in Fig.~\ref{fig:setup_led_panel}(a).
The prototype consists of an LED panel transmitter and an event-camera receiver. The panel contains eight individually addressable LEDs arranged in a $2\times4$ grid. The receiver is a LUCID Triton2 EVS event camera, positioned facing the panel under a line-of-sight (LoS) condition. The event camera reports asynchronous brightness transitions induced by LED switching with microsecond-level timestamps.
Despite their high temporal resolution, event cameras exhibit non-idealities, including pixel front-end dynamics, non-instantaneous response, trailing effects, readout delay, and refractory behavior, leading to timestamp inconsistency~\cite{ECO-COMM}. Following the mitigation strategies in~\cite{ECO-COMM}, we avoid relying on single-pixel responses and instead aggregate events over a small region of interest (ROI) for each LED. Specifically, we define a $9\times9$ pixel window centered at each LED’s nominal image location and use this ROI for event aggregation and detection (illustrated as dark squares in Fig.~\ref{fig:setup_led_panel}(c)). The ROI locations and LED indexing are determined after device detection via the localization procedure described in Section~\ref{sec:protocol-single}.

The system supports $U$ users operating asynchronously. User $u\in\{1,2,\dots,U\}$ is assigned a dedicated subset of LEDs $\mathcal{L}_u\subseteq\mathcal{L}$, and the subsets are non-overlapping, i.e., $
\mathcal{L}_u \cap \mathcal{L}_v=\emptyset, \quad \forall u\neq v$.
This spatial partitioning emulates practical deployments where users transmit from spatially distinct sources. Notably, our prototype represents a conservative (worst-case) setting in which user regions are adjacent on the same panel, thereby minimizing spatial separation. As shown in Fig.~\ref{fig:setup_led_panel}(b), we partition the panel into two regions: the left four LEDs are assigned to User~1 (UE1) and the right four LEDs are assigned to User~2 (UE2).
ECO-ID operates without active focusing (the lens focus is fixed at infinity). As a result, LEDs appear spatially blurred in the event stream (Fig.~\ref{fig:setup_led_panel}(c)). Empirically, maintaining perfect focus does not provide a consistent performance benefit for ECO-ID. Consequently, the system avoids autofocus-induced latency.

Time is discretized into slots of duration $T=1/f$, where $f$ is the LED switching (symbol) frequency. Users transmit asynchronously: user $u$ begins at an arbitrary time offset $\tau_u$, which may differ across users. Let $x_\ell(t)\in\{0,1\}$ denote the instantaneous state of LED $\ell$ at time $t$, and define the set of active LEDs as $\mathcal{A}(t) \triangleq \{\ell\in\mathcal{L}: x_\ell(t)=1\}$.
When multiple users transmit concurrently, the received optical pattern corresponds to the union of active LEDs across users. Since the assigned LED sets are disjoint, concurrent transmissions superpose in intensity but remain separable by their spatial association to distinct ROIs.

The receiver outputs an asynchronous stream of events $e_i=(r_i,c_i,t_i,p_i)$, where $(r_i,c_i)$ is the pixel location, $t_i$ is the timestamp, and $p_i\in\{+1,-1\}$ denotes polarity (increase/decrease in log-intensity). For each LED $\ell$, we define an ROI $\mathcal{R}_\ell$ and extract per-LED activity by aggregating events within $\mathcal{R}_\ell$. For example, the event count over an interval $[t,t+\delta)$ is
\begin{equation}
N_\ell([t,t+\delta)) \triangleq 
\left|\left\{e_i : (r_i,c_i)\in\mathcal{R}_\ell,\; t_i \in [t,t+\delta)\right\}\right|.
\end{equation}
The collection of traces $\{N_\ell(\cdot)\}_{\ell\in\mathcal{L}}$ constitutes the receiver-side observation used for subsequent detection and decoding.

Without loss of generality, we assume a LoS optical channel between the transmitter panels and the receiver. The event camera’s asynchronous output enables precise estimation of LED switching instants under asynchronous operation, while disjoint LED allocation provides scalable multi-user separability without conventional multiple-access scheduling.

\section{ECO-ID: Multi-User Ultra-Low Latency Identification}\label{sec:ECO-ID}

\subsection{Overview}

ECO-ID defines a signaling structure that exploits both \emph{spatial separation} (disjoint LED assignment) and \emph{temporal encoding} (user-specific timing) to enable concurrent multi-user identification without inter-user synchronization. For each user $u$, the assigned LED set $\mathcal{L}_u$ is partitioned into a localization subset $\mathcal{L}^{\mathrm{loc}}_u \subset \mathcal{L}_u$ and an identification LED $\ell^{\mathrm{id}}_u \in \mathcal{L}_u \setminus \mathcal{L}^{\mathrm{loc}}_u$. The localization LEDs indicate transmission onset, while the identification LED encodes the user identity via a user-specific delay.
User $u$ initiates its identification sequence at time $\tau_u$ by switching the LEDs in $\mathcal{L}^{\mathrm{loc}}_u$ (near-)simultaneously. After a delay $T^{(u)}_{\mathrm{ID}}$ relative to this onset, user $u$ switches $\ell^{\mathrm{id}}_u$. Any remaining slots can be used for payload transmission (if required) under the chosen modulation format.
At the receiver, the start time is estimated by detecting a coincident burst of activity in the ROIs corresponding to $\mathcal{L}^{\mathrm{loc}}_u$, yielding $\widehat{t}^{(u)}_{\mathrm{start}}$. The identification LED activation time is similarly detected as $\widehat{t}^{(u)}_{\mathrm{ID}}$, and the estimated ID delay is
\begin{equation}
\widehat{T}^{(u)}_{\mathrm{ID}} \triangleq \widehat{t}^{(u)}_{\mathrm{ID}} - \widehat{t}^{(u)}_{\mathrm{start}}.
\end{equation}
The user identity is obtained by mapping $\widehat{T}^{(u)}_{\mathrm{ID}}$ to the closest entry in a predefined timing codebook. After establishing $\widehat{t}^{(u)}_{\mathrm{start}}$ as the timing reference, the payload (if present) is decoded by sampling per-LED activity traces $\{N_\ell(\cdot)\}$ in symbol windows of duration $T$ (or integer multiples of $T$) aligned to $\widehat{t}^{(u)}_{\mathrm{start}}$.
ECO-ID therefore separates users spatially through disjoint LED sets and distinguishes identities temporally through user-specific ID delays, leveraging the event camera’s microsecond-level timestamps to support fully asynchronous multi-user identification. After device identification, higher-layer authentication and key establishment (e.g., a standard 4-way handshake) can be executed as needed by the target system and threat model.

\subsection{Single-User Identification Protocol}\label{sec:protocol-single}

\begin{figure}[ht!]
\centering
\includegraphics[width=.95\columnwidth]{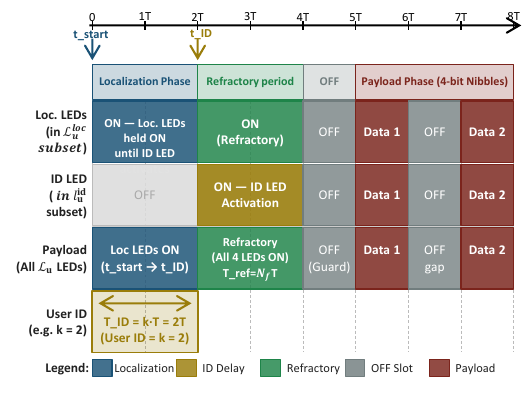}
\caption{Timing diagram for single-UE identification protocol.}
\label{fig:timing diagram_single}
\end{figure}

In the single-user setting, one user equipment (UE) is assigned four LEDs on an LED panel. Transmission is organized into discrete time slots of duration $T = 1/f$, where $f$ denotes the LED switching (symbol) frequency. The UE transmits an identification frame consisting of the sequential phases below. The receiver is an event camera that infers LED switching instants by monitoring event activity within predefined LED-specific ROIs. Fig.~\ref{fig:timing diagram_single} depicts a timing diagram for the single-UE identification protocol, described in four phases below.

\textbf{(i) Localization.} At the beginning of the frame, a predefined set of three LEDs is switched \emph{ON} simultaneously and held \emph{ON} for $N_{\mathrm{loc}}$ slots (a fixed design parameter). This multi-LED preamble produces a coincident burst of events in the corresponding ROIs, enabling robust device detection and LED indexing without global synchronization. Let $\widehat{t}_{\mathrm{start}}$ denote the receiver-estimated onset time.

\textbf{(ii) Identification.} After localization, a fourth LED (reserved for identification) is switched \emph{ON} after a delay
\begin{equation}
T_{\mathrm{ID}} = kT,
\end{equation}
where $k \in \mathbb{Z}_{\ge 0}$ uniquely defines a user. Let $\widehat{t}_{\mathrm{ID}}$ be the detected activation time of the identification LED. The decoded timing code is obtained by quantizing the measured delay to the nearest slot $\widehat{k} = \mathrm{round}\!\left(({\widehat{t}_{\mathrm{ID}}-\widehat{t}_{\mathrm{start}}})/{T}\right) = \mathrm{round}\!\left(\widehat{T}_{\mathrm{ID}}/{T}\right)$,
equivalently by nearest-neighbor matching to the valid set of timing codes.

\textbf{(iii) Refractory/guard and delimiter.} Following identification, all four LEDs remain \emph{ON} for a refractory duration $
T_{\mathrm{ref}} = N_f T$,
where $N_f$ is a design parameter. This guard interval mitigates transient effects and spurious events after rapid switching by providing a stable brightness level. The frame then inserts a one-slot delimiter by switching all LEDs \emph{OFF} for one slot, creating a clear temporal boundary between the preamble (localization/identification) and the payload and producing a strong transition that aids timing re-alignment.

\textbf{(iv) Payload transmission (optional).} If a payload is transmitted, the four LEDs form a 4-bit symbol alphabet. In each payload \emph{data} slot, the LED state vector $\mathbf{s}[n]\in\{0,1\}^4$ encodes one nibble $b[n]\in\{0,\dots,15\}$ using a fixed bit-to-LED mapping. To ensure reliable event generation, each data slot is followed by one \emph{OFF} slot (all LEDs OFF). Consequently, one byte requires two nibbles and four slots in total (data-OFF-data-OFF). The receiver decodes the payload by aggregating events within each ROI over decision windows aligned to the symbol boundaries and inferring $\mathbf{s}[n]$ from the observed per-LED activity.

\subsection{Multi-User Identification Protocol}

The proposed protocol extends naturally to the multi-user setting without requiring inter-user synchronization or explicit coordination. This is enabled by two orthogonal mechanisms: (i) \emph{spatial separation} through disjoint LED assignments and (ii) \emph{temporal identification} through user-specific delay signatures. As a result, multiple UEs can initiate identification frames asynchronously and may partially or fully overlap in time while remaining separable at the receiver. In our prototype, multi-user operation is emulated by logically partitioning a single $2\times4$ LED panel into two independent regions, corresponding to two adjacent user transmitters. Specifically, UE1 uses LEDs $\{0,1,4,5\}$ and UE2 uses LEDs $\{2,3,6,7\}$. Because these subsets are disjoint, events generated by one UE are confined to the ROIs of its assigned LEDs and do not contaminate the ROIs of other UEs, providing a first layer of separation. More generally, any disjoint LED subsets can be assigned to users, and different temporal IDs can be realized within each subset.

\begin{figure}[ht!]
\centering
\includegraphics[width=.95\linewidth]{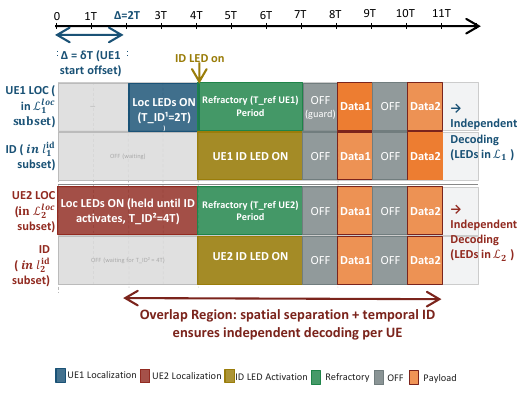}
\caption{Timing diagram for multi-UE identification protocol.}
\label{fig:timing diagram-multi}
\end{figure}

As depicted in Fig.~\ref{fig:timing diagram-multi} users transmit asynchronously. Let UE2 start at $t=0$ and UE1 start at $t=\Delta$, where $\Delta=\delta T$ and $\delta\in\mathbb{Z}_{\ge 0}$. The receiver does not require global alignment across users; instead, each UE is time-referenced to its own detected localization onset, so overlap does not prevent decoding.
Within each user’s assigned LED subset, identity is encoded temporally via the delay between localization and the identification LED. Concretely, user $u$ employs $T^{(u)}_{\mathrm{ID}} = k_u T$,
where $k_u$ is an integer timing code. For example, UE1 may use $T^{(1)}_{\mathrm{ID}}=2T$ and UE2 may use $T^{(2)}_{\mathrm{ID}}=4T$. The receiver estimates $T^{(u)}_{\mathrm{ID}}$ independently from the event activity within the ROIs corresponding to $\mathcal{L}_u$; therefore, concurrent transmissions do not interfere with identification as long as ROIs remain distinct.
At any time slot, the emitted light pattern equals the union of active LEDs across all UEs (implemented as a bitwise OR across user LED state vectors). Since each UE controls a disjoint LED subset, this optical superposition introduces no ambiguity: activity observed within a given ROI can be attributed to exactly one UE.

Decoding is performed independently per UE using only the ROIs associated with its LEDs. For each user $u$, the receiver (i) detects frame onset from coincident activity in the localization-LED ROIs $\mathcal{L}^{\mathrm{loc}}_u$, (ii) infers identity from the measured delay to the identification LED activation, and (iii) decodes the payload (if transmitted) using only the assigned LEDs. Consequently, even with overlapping frames, each UE can be localized, identified, and decoded without scheduling, coordination, or synchronization among users.

\subsection{Analysis on Maximum User ID Support}

\begin{table}[b]
\centering
\caption{Frequency-dependent ID-slot capacity and assignment combinations (1\,ms window).}
\label{tab:id_slot_capacity}

\setlength{\tabcolsep}{4pt}
\renewcommand{\arraystretch}{1.05}
\tiny
\resizebox{\columnwidth}{!}{%
\begin{tabular}{ccccccc}
\toprule
{$f$ (Hz)} & {$T$ ($\mu$s)} & \textbf{$N$} & \textbf{$U_{\max}$} &
\textbf{$P(N,2)$} & \textbf{$P(N,3)$} & \textbf{$P(N,4)$} \\
\midrule
4000  & 250.0 & 4  & 4  & 12  & 24   & 24    \\
6000  & 166.7 & 6  & 6  & 30  & 120  & 360   \\
8000  & 125.0 & 8  & 8  & 56  & 336  & 1680  \\
10000 & 100.0 & 10 & 10 & 90  & 720  & 5040  \\
12000 & 83.3  & 12 & 12 & 132 & 1320 & 11880 \\
14000 & 71.4  & 14 & 14 & 182 & 2184 & 24024 \\
16000 & 62.5  & 16 & 16 & 240 & 3360 & 43680 \\
18000 & 55.6  & 18 & 18 & 306 & 4896 & 73440 \\
20000 & 50.0  & 20 & 20 & 380 & 6840 & 116280 \\
\bottomrule
\end{tabular}%
}
\end{table}

We characterize the maximum number of simultaneously active users supported by ECO-ID as a function of the LED switching frequency. The analysis follows ECO-ID’s temporal identification mechanism, in which each user encodes its identity via the delay between the localization onset and the activation of its identification LED.
Let $f$ (Hz) denote the switching frequency, yielding a slot duration $T={1}/{f}$.
We restrict the identification delay to a fixed window of length $T_{\mathrm{win}}=1$~ms. The number of distinct discrete delay slots within this window is
\begin{equation}
N=\left\lfloor{T_{\mathrm{win}}}/{T}\right\rfloor
=\left\lfloor{f}/{1000}\right\rfloor.
\end{equation}
To avoid ambiguity, simultaneously active users must select distinct delay slots; i.e., no two users may use the same identification delay.
In addition to temporal resolution, multi-user capacity is constrained by spatial resources. Let $K$ denote the number of independent spatial regions available for user separation (e.g., distinct panels or disjoint LED groups). The maximum number of simultaneous users is therefore
\begin{equation}
U_{\max}=\min(K,N),
\end{equation}
capturing the joint limitation imposed by spatial separation and temporal slot availability.

For a given number of active users $U\le N$, the number of possible \emph{ordered} assignments of unique ID slots is
\begin{equation}
P(N,U)={N!}/{(N-U)!},
\end{equation}
which counts the distinct ways to map $U$ users to $U$ different delay slots. For example, at $f=4000$~Hz, $T=250~\mu$s and $N=4$ slots fit within the 1~ms window. If spatial resources are sufficient, up to four users can be supported, and the ordered assignments are $P(4,2)=12$, $P(4,3)=24$, and $P(4,4)=24$ for $U=2,3,4$, respectively.

Table~\ref{tab:id_slot_capacity} summarizes $T$, $N$, and representative values of $P(N,U)$ across frequencies.
As $f$ increases, $T$ decreases and $N$ increases, enabling more distinct temporal IDs within the same window. In practice, however, $K$ can be the tighter bottleneck: in our prototype with $K=2$, at most two users can be active simultaneously regardless of $N$ (provided $N\ge2$). Nevertheless, larger $N$ increases assignment flexibility and reduces the probability of ID-slot collisions in unscheduled multi-user operation.
Although the number of users that can be read simultaneously is ultimately bounded by the event camera’s spatial resolution, the available pixel budget is still substantial. For example, with a $1920\times720$ sensor, one can accommodate on the order of $10^4$ non-overlapping $9\times9$ ROIs, which corresponds to $\mathcal{O}(10^3)$ users when allocating an 8-LED panel (or equivalently, an 8-ROI region) per user. In practice, the supported user count will be lower due to optical blur, inter-LED crosstalk, ROI overlap, and geometric constraints on placing many light sources within the field of view. Nevertheless, this analysis highlights a clear scaling advantage over existing identification technologies. Overall, ECO-ID scales by combining spatial separation with temporal encoding, enabling concurrent multi-user identification without synchronization or explicit resource scheduling.

\section{Evaluation}\label{sec:evaluation}

Using the prototype system described in Section~\ref{sec:system model}, we evaluate ECO-ID with a focus on \emph{multi-user capability} and \emph{ultra-low-latency} device identification. Details on event-camera configuration, parameter optimization, and achievable end-to-end payload latency are provided in our prior work~\cite{ECO-COMM}. This section concentrates on the identification pipeline.

\subsection{Localization and User Identification}

\begin{figure*}[ht!]
    \centering
    \begin{subfigure}[t]{0.32\textwidth}
        \centering
        \includegraphics[width=\linewidth]{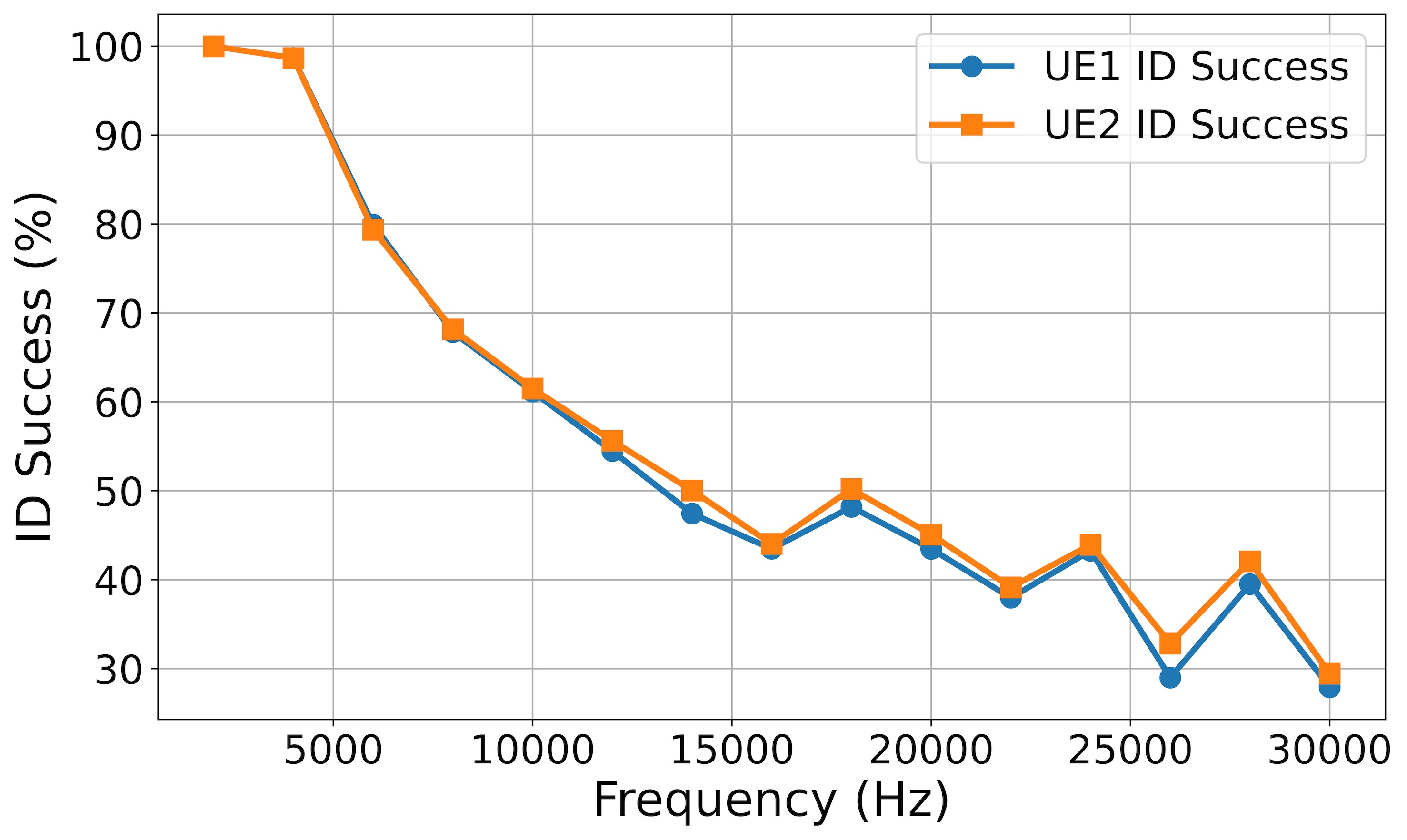}
        \caption{ID success rate vs. frequency.}
        \label{fig:id_success}
    \end{subfigure}\hfill
    \begin{subfigure}[t]{0.32\textwidth}
        \centering
        \includegraphics[width=\linewidth]{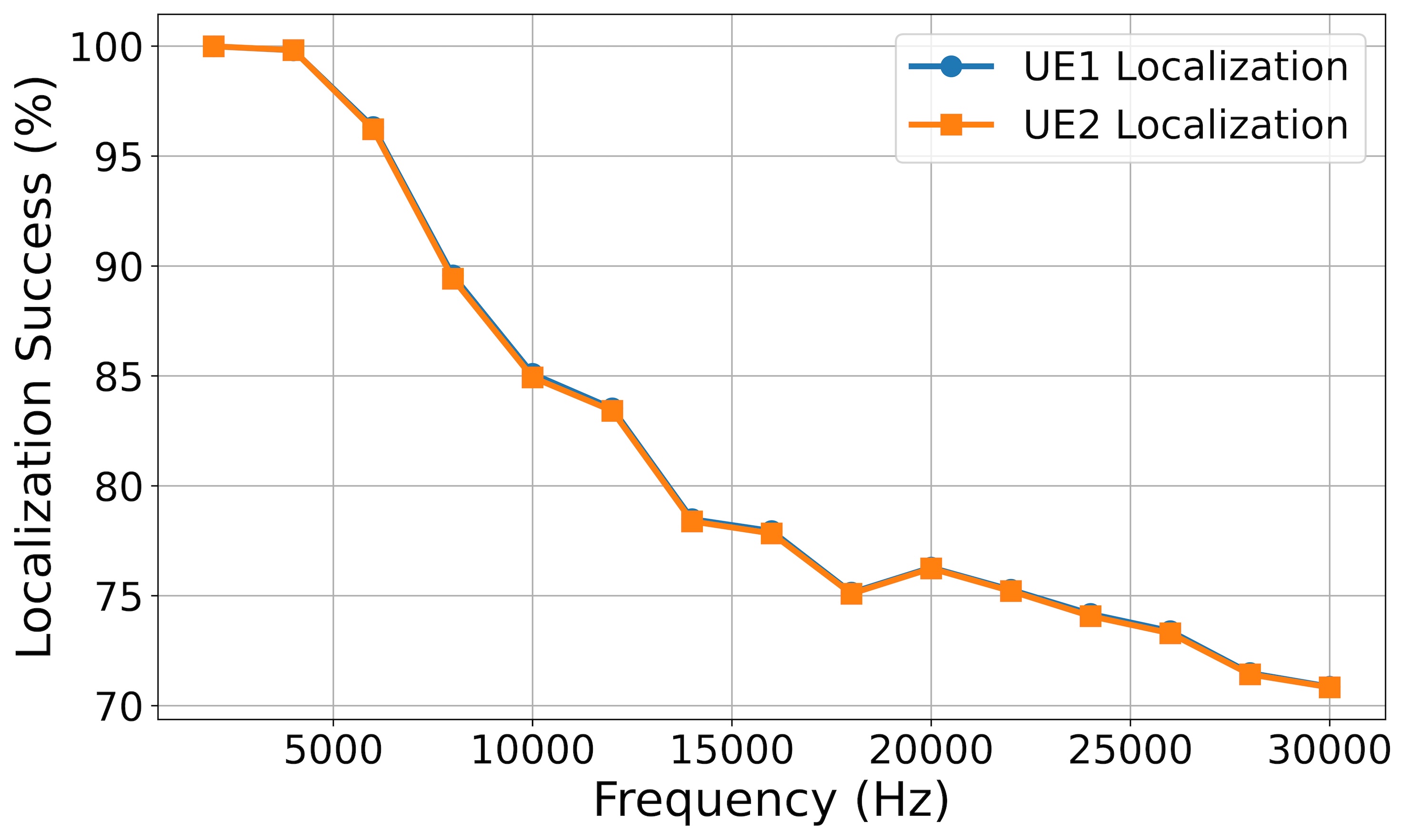}
        \caption{Localization success rate vs. frequency.}
        \label{fig:localization_success}
    \end{subfigure}\hfill
    \begin{subfigure}[t]{0.32\textwidth}
        \centering
        \includegraphics[width=\linewidth]{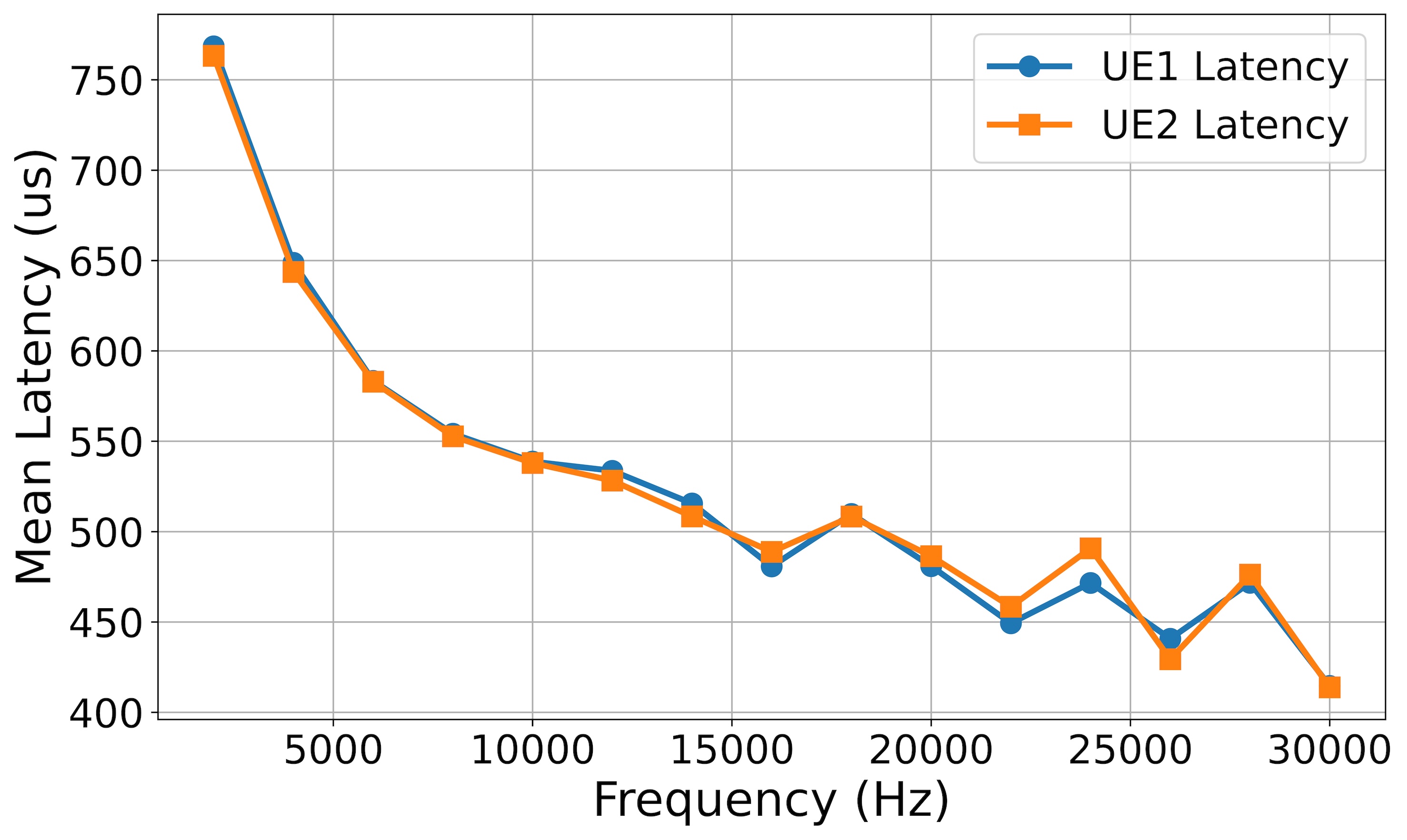}
        \caption{Mean ID latency vs. frequency.}
        \label{fig:latency_plot}
    \end{subfigure}
    \caption{Evaluation results on two-user identification.}
    \label{fig:two_user_results}
\end{figure*}

We evaluate the localization and user-identification phases over LED switching frequencies from $2$~kHz to $30$~kHz. Unless otherwise stated, the transmitter--receiver distance is fixed at $1$~m. For each frequency, we run $10{,}000$ independent trials. In each trial, the system randomly selects a pair of identification delays $(T^{(1)}_{\mathrm{ID}},T^{(2)}_{\mathrm{ID}})$ for UE1 and UE2, respectively, with $T^{(1)}_{\mathrm{ID}}\neq T^{(2)}_{\mathrm{ID}}$ to enforce distinct temporal IDs. Fig.~\ref{fig:id_success} shows the identification success rate versus frequency. Identification remains near $100\%$ at low frequencies (2-4~kHz) and degrades as frequency increases, as higher switching rates reduce the temporal separation between transitions and make timing estimation more susceptible to missed events and sensor non-idealities. Fig.~\ref{fig:localization_success} reports localization success, which follows a similar but less pronounced trend: localization remains high at low frequencies and gradually decreases at higher frequencies, primarily because shorter ON intervals generate fewer events per symbol and reduce detection robustness. Fig.~\ref{fig:latency_plot} plots mean identification latency versus frequency. As expected, latency decreases with frequency due to shorter symbol durations, but this improvement comes at the cost of reduced reliability.

\begin{figure*}[ht!]
\centering
\begin{subfigure}{0.19\linewidth}
\includegraphics[width=\linewidth]{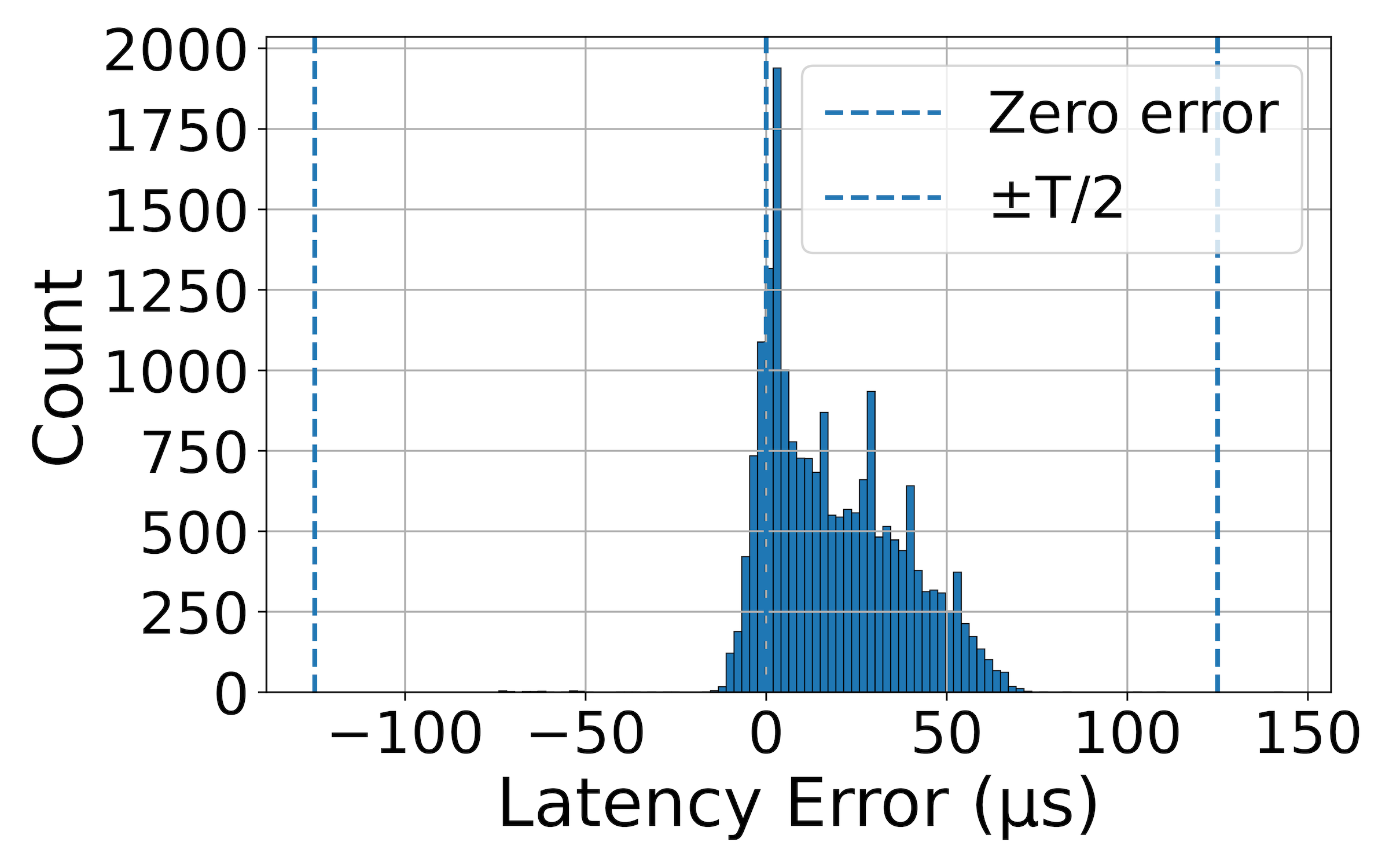}
\caption{4000 Hz}
\end{subfigure}
\begin{subfigure}{0.19\linewidth}
\includegraphics[width=\linewidth]{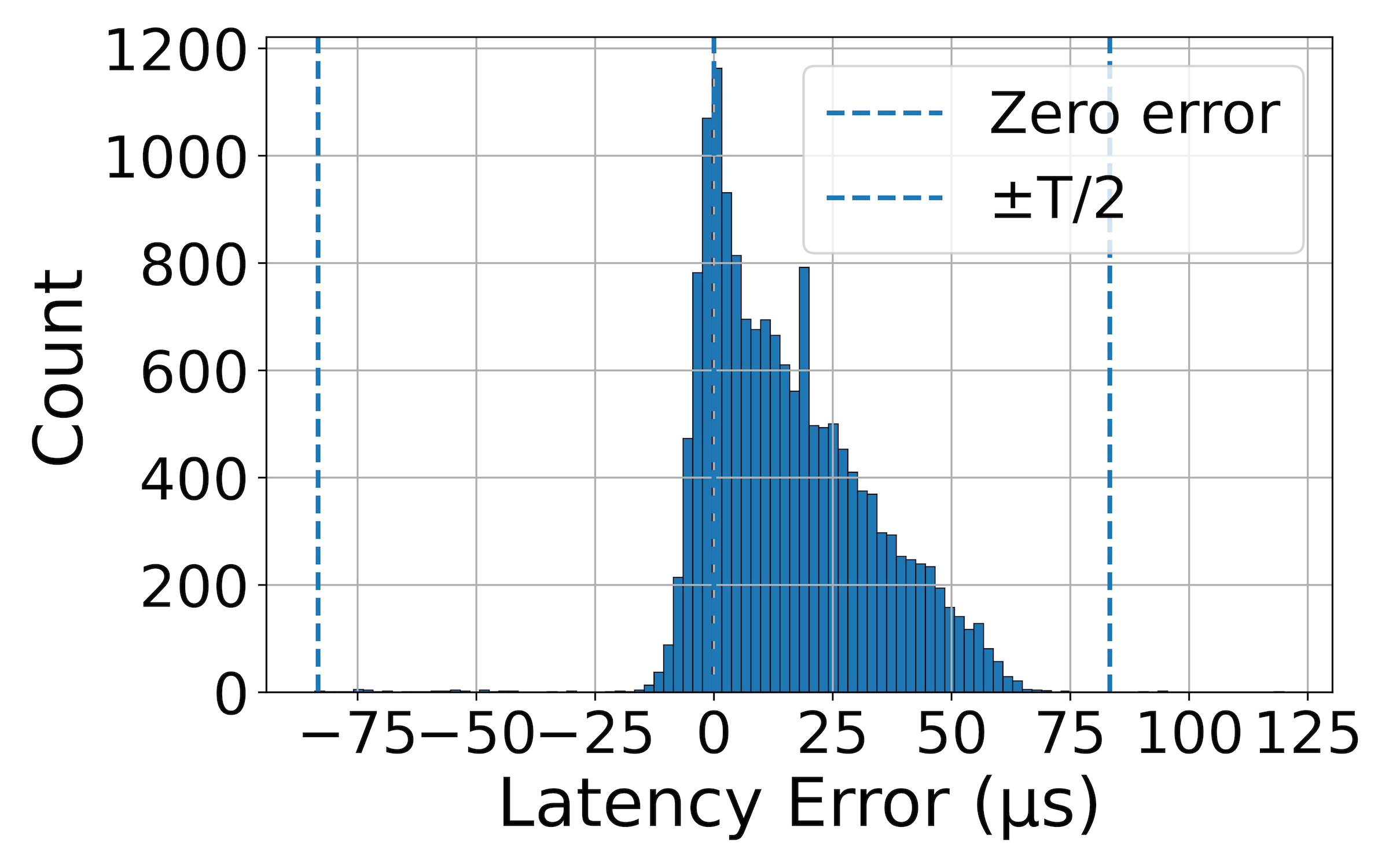}
\caption{6000 Hz}
\end{subfigure}
\begin{subfigure}{0.19\linewidth}
\includegraphics[width=\linewidth]{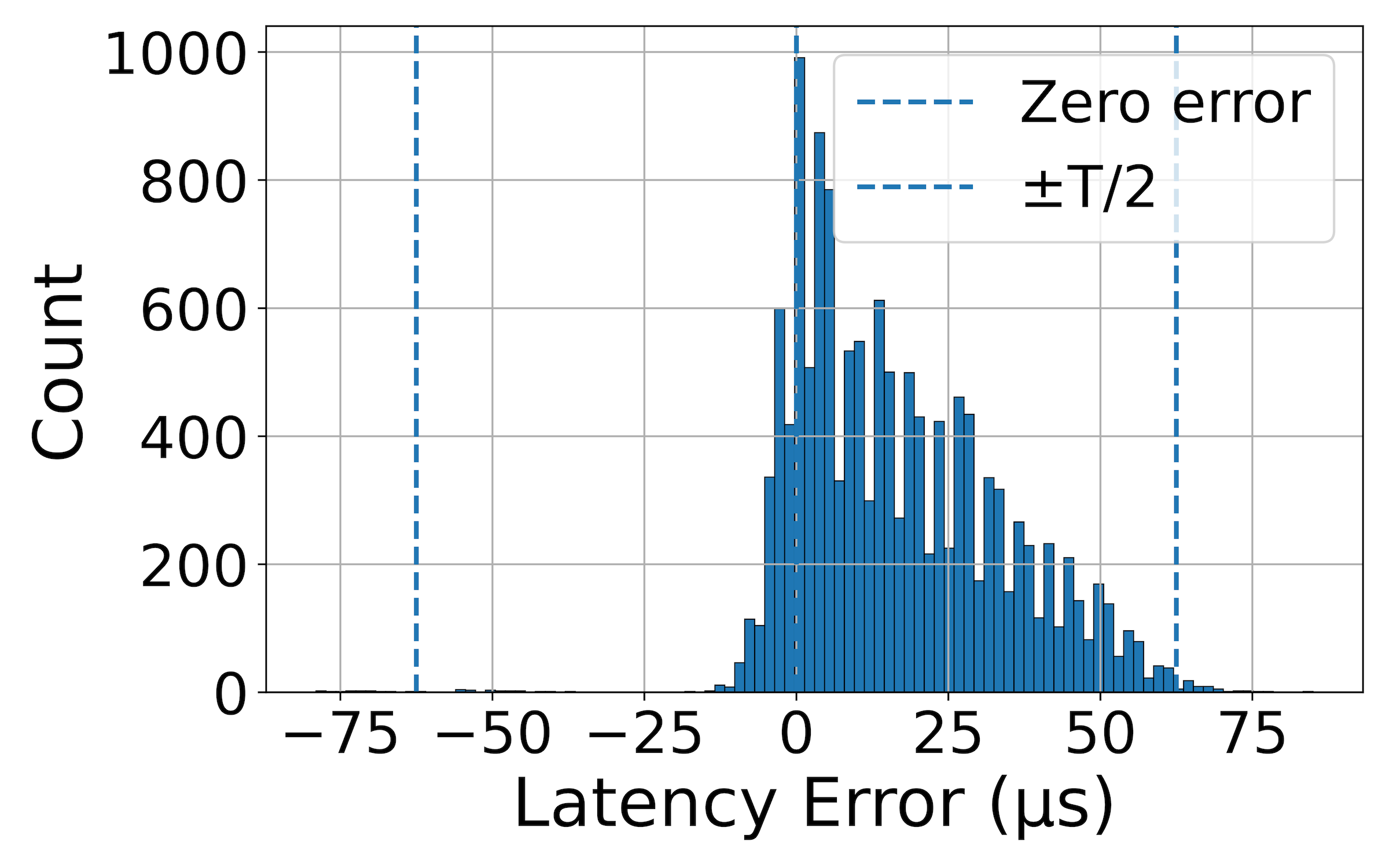}
\caption{8000 Hz}
\end{subfigure}
\begin{subfigure}{0.19\linewidth}
\includegraphics[width=\linewidth]{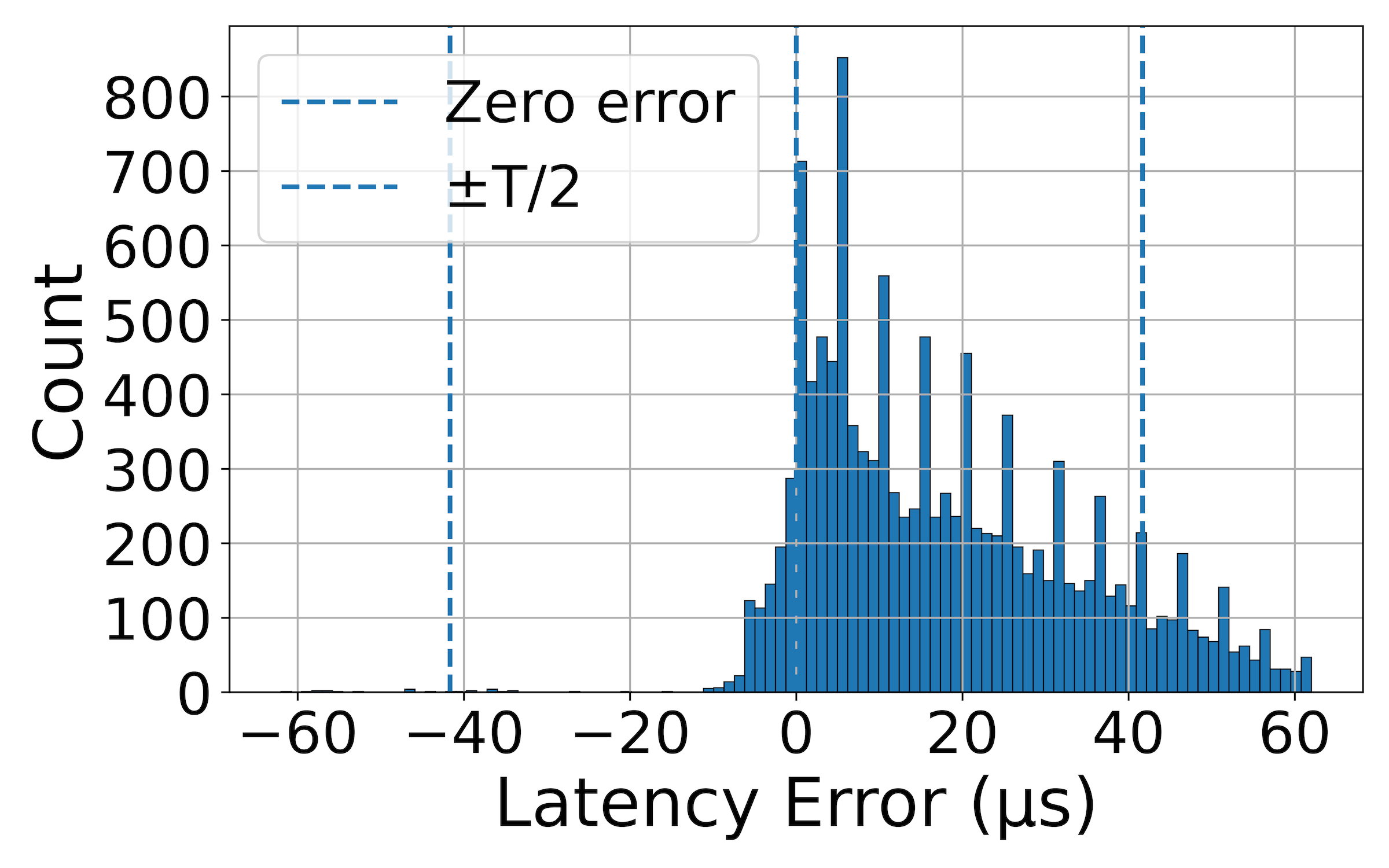}
\caption{12000 Hz}
\end{subfigure}
\begin{subfigure}{0.19\linewidth}
\includegraphics[width=\linewidth]{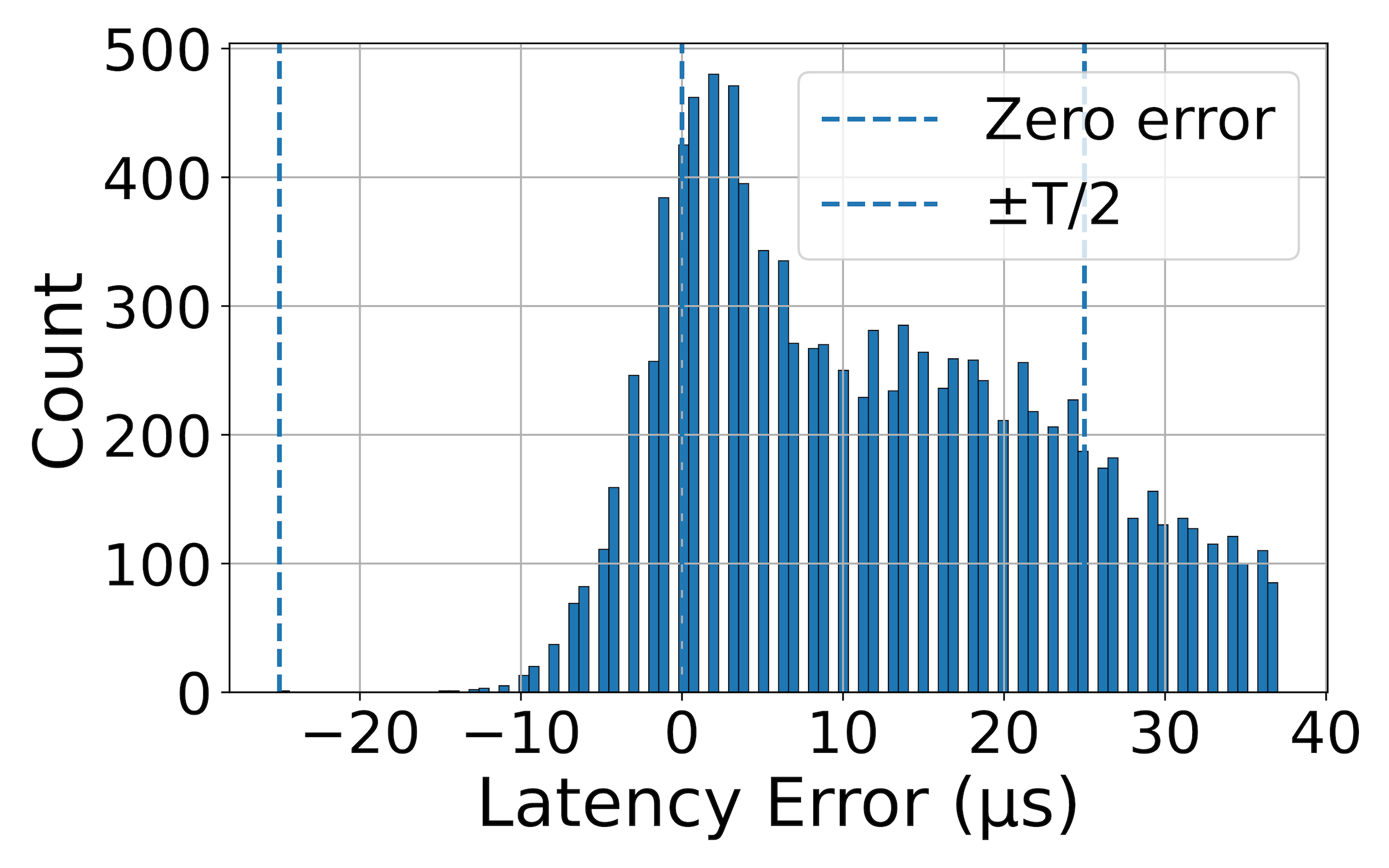}
\caption{20000 Hz}
\end{subfigure}
\caption{Latency error histograms at different frequencies. The dashed lines represent the $\pm T/2$ decision boundaries for correct ID detection.}
\label{fig:latency_histograms}
\end{figure*}

\begin{table}[ht!]
\centering
\caption{Localization, identification, and latency performance across frequencies.}
\label{tab:performance_table}
\setlength{\tabcolsep}{3pt}        % 
\renewcommand{\arraystretch}{1.05} 
\footnotesize                       
\resizebox{\columnwidth}{!}{
\begin{tabular}{cccccccc}
\toprule
\textbf{Freq} & \textbf{Trials} & \textbf{UE1 Loc} & \textbf{UE2 Loc} & \textbf{UE1 ID} & \textbf{UE2 ID} & \textbf{UE1 Lat} & \textbf{UE2 Lat} \\
\textbf{(Hz)} &                & \textbf{(\%)}    & \textbf{(\%)}    & \textbf{(\%)}   & \textbf{(\%)}   & \textbf{($\mu$s)}& \textbf{($\mu$s)}\\
\midrule
2000  & 10000 & 99.99 & 99.99 & 99.98 & 99.98 & 768.51 & 763.27 \\
4000  & 10000 & 99.81 & 99.82 & 98.67 & 98.66 & 648.77 & 643.70 \\
6000  & 10000 & 96.32 & 96.22 & 79.93 & 79.34 & 583.44 & 582.94 \\
8000  & 10000 & 89.57 & 89.42 & 67.85 & 68.15 & 554.19 & 552.72 \\
10000 & 10000 & 85.09 & 84.93 & 61.16 & 61.50 & 538.74 & 537.98 \\
12000 & 10000 & 83.52 & 83.41 & 54.46 & 55.63 & 533.66 & 528.22 \\
14000 & 10000 & 78.48 & 78.38 & 47.43 & 50.02 & 515.60 & 508.40 \\
16000 & 10000 & 77.94 & 77.83 & 43.47 & 44.02 & 480.73 & 488.81 \\
18000 & 10000 & 75.14 & 75.09 & 48.18 & 50.19 & 509.77 & 508.33 \\
20000 & 10000 & 76.28 & 76.24 & 43.46 & 45.08 & 480.89 & 486.00 \\
22000 & 10000 & 75.27 & 75.21 & 37.97 & 39.14 & 449.28 & 458.42 \\
24000 & 10000 & 74.17 & 74.07 & 43.25 & 43.93 & 471.57 & 490.74 \\
26000 & 10000 & 73.39 & 73.29 & 28.98 & 32.82 & 440.73 & 429.37 \\
28000 & 10000 & 71.48 & 71.42 & 39.50 & 42.06 & 471.70 & 476.19 \\
30000 & 10000 & 70.86 & 70.83 & 27.89 & 29.44 & 414.64 & 413.82 \\
\bottomrule
\end{tabular}%
}
\end{table}

Table~\ref{tab:performance_table} consolidates these results and highlights the reliability-latency trade-off. At $2$~kHz, ECO-ID achieves $99.99\%$ localization and $99.98\%$ identification with $\sim$0.76~ms mean latency, while at $4$~kHz it maintains $\approx 99.8\%$ localization and $\approx 98.7\%$ identification with a reduced latency of $\sim$0.64~ms. Beyond $4$~kHz, identification degrades substantially faster than localization (e.g., at $6$~kHz localization remains above $96\%$ but identification drops to $\sim 79\%$; at $10$~kHz localization is still $\sim 85\%$ while identification falls to $\sim 61\%$), consistent with the fact that localization relies on a strong coincident event burst across multiple LEDs, whereas identification requires precise relative-delay estimation under fewer events per symbol. Latency continues to decrease overall (from $\sim$768~$\mu$s at $2$~kHz to $\sim$415~$\mu$s at $30$~kHz), with minor non-monotonicities at high frequencies largely due to averaging over successful trials. Across all frequencies, UE1 and UE2 exhibit nearly symmetric performance, indicating that the adjacent spatial partitioning in this conservative setup does not introduce meaningful user bias.

\subsection{Latency Error Analysis for Identification}

To explain the degradation in localization and identification at higher switching frequencies, we examine the distribution of \emph{identification-delay estimation errors} using latency-error histograms. For each trial, the latency error is $
\varepsilon \triangleq \widehat{T}^{(u)}_{\mathrm{ID}} - T^{(u)}_{\mathrm{ID}}$,
where $T^{(u)}_{\mathrm{ID}}=k_u T$ is the ground-truth identification delay (an integer multiple of the slot duration $T$) and $\widehat{T}^{(u)}_{\mathrm{ID}}$ is the delay estimated from the event stream. As described in Section~\ref{sec:ECO-ID}, decoding maps $\widehat{T}^{(u)}_{\mathrm{ID}}$ to the nearest discrete ID slot. Therefore, correct identification occurs if and only if the estimation error falls within the nearest-neighbor decision region $
\varepsilon \in \left(-T/2,\,T/2\right)$,
constituting the decision boundary. When $|\varepsilon|>T/2$, the measured delay is quantized to an adjacent slot, resulting in an incorrect user ID.
Fig.~\ref{fig:latency_histograms} shows the latency-error distributions at representative frequencies. At $4$~kHz ($T=250~\mu$s), errors are tightly concentrated around zero and lie well within the $\pm T/2$ boundary, consistent with the near-perfect identification accuracy observed in Fig.~\ref{fig:id_success} and Table~\ref{tab:performance_table}. At $6$~kHz ($T\approx 166.7~\mu$s), the distribution widens slightly but remains largely within the decision region, leading to a moderate reduction in success rate. At $8$~kHz ($T=125~\mu$s), a non-negligible fraction of errors crosses $\pm T/2$, increasing the probability of slot misassignment and producing the noticeable drop in identification accuracy. At higher frequencies (e.g., $12$~kHz and $20$~kHz, with $T=83.3~\mu$s and $50~\mu$s), the allowable error margin shrinks further and a substantial portion of the distribution falls outside the decision boundary, causing frequent misclassification and pronounced performance degradation.
This analysis indicates that the observed degradation at high switching frequencies is fundamentally driven by reduced temporal tolerance: increasing $f$ lowers the mean latency but simultaneously tightens the quantization margin, making identification more sensitive to event jitter, missed transitions, and detection uncertainty.

% \subsection{Discussion}

Overall, the evaluation results demonstrate that the proposed ECO-ID enables simultaneous, low-latency identification without inter-user synchronization or explicit coordination. By assigning disjoint LED subsets to different users (spatial separation) and encoding identities in user-specific delay signatures (temporal coding), the receiver can localize, associate, and identify multiple users independently, even when transmissions overlap, without relying on conventional multiple-access scheduling such as TDMA/FDMA. The event-driven sensing pipeline further reduces overhead and latency by reacting only to brightness transitions, yielding sub-millisecond identification while remaining robust without focusing. Taken together, these findings validate ECO-ID as a scalable and flexible multi-user identification approach that supports asynchronous operation with minimal coordination overhead. 

\section{conclusion}

In this paper, we presented ECO-ID, an event-camera based identification and authentication system for time-critical, multi-user interactive environments. ECO-ID combines spatial separation via disjoint LED assignments with temporal identity encoding through user-specific delay signatures, enabling concurrent, asynchronous identification without explicit scheduling or coordination. Using ROI-based event aggregation, the receiver robustly detects localization and identification transitions even without focusing, while event-driven sensing achieves sub-millisecond end-to-end identification latency. A prototype implementation validates ECO-ID and reveals a reliability-latency trade-off. Our current setup practically delivers $\sim$99.8\% localization, $\sim$98.7\% identification, and $\sim$0.64~ms mean latency for two users. Future work will extend scaling to more users and panels, strengthen timing-robust decoding under motion and ambient light, and evaluate ECO-ID in larger interactive deployments.

\bibliographystyle{IEEEtran}
\bibliography{references}
\end{document}